\documentclass[12pt]{article}
\usepackage[dvips]{graphics,color}
\usepackage{bbold}
\usepackage{epsfig}
\def\bea{\begin{eqnarray}}
\def\eea{\end{eqnarray}}
\def\be{\begin{equation}}
\def\ee{\end{equation}}
\def\nn{\nonumber}
\def\le{\left}
\def\re{\right}

\def\i{i }
\newcommand{\ft}[2]{{\textstyle\frac{#1}{#2}}\,}

\def\a{\alpha}
\def\b{\beta}

\def\d{\delta}

\def\e{\epsilon}

\def\g{\gamma}
\def\G{\Gamma}

\def\m{\mu}
\def\n{\nu}

\def\r{\rho}

\def\t{\tau}
\def\th{\theta}
\def\Th{\Theta}

\def\tS{\tilde S}

\newcommand{\bX}{\hat{\bf X}}

\newcommand{\eqn}[1]{(\ref{#1})}
\newcommand{\XX}[2]{\{X^#1,X^#2\}}
\newcommand{\NPB}[3]{{Nucl.\ Phys.} {\bf B#1} (#2) #3}
\newcommand{\IJMPA}[3]{{Int.\ J.\ Mod.\ Phys.} {\bf A#1} (#2) #3}
\newcommand{\CMP}[3]{{Commun.\ Math.\ Phys.} {\bf #1} (#2) #3}
\newcommand{\PRD}[3]{{Phys.\ Rev.} {\bf D#1} (#2) #3}
\newcommand{\PLB}[3]{{Phys.\ Lett.} {\bf B#1} (#2) #3}
\newcommand{\JHEP}[3]{{JHEP} {\bf #1} (#2) #3}

\newcommand{\CQG}[3]{{Class.\ Quant.\ Grav.} {\bf #1} (#2) #3}
\newcommand{\T}[1]{``{\em #1}''}
\newcommand{\MPLA}[3]{{Mod.\ Phys.\ Lett.}{\bf A#1}(#2) #3}
\newcommand{\hepth}[1]{{\tt hep-th/#1}}

\begin{document}
\begin{flushright}
{\small AEI-2001-064}\\[3mm]
\end{flushright}

\vspace{1cm}

\begin{center}
{\Large An Introduction to the Quantum Supermembrane} {\footnote {Lectures 
given by Hermann Nicolai at the {\it 4th Mexican School on
Gravitation and Mathematical Physics}, 3-8 December 2000, Huatulco, Mexico.}}\\

\vspace{2cm}

{\it Arundhati Dasgupta, Hermann Nicolai and Jan Plefka}\\

\vspace{1cm}

{ Max-Planck-Institut f\"ur Gravitationsphysik, \\
 Albert-Einstein-Institut,\\
 Golm, Germany}
\end{center}

\vspace{1cm}

{\bf Abstract:} We review aspects of quantisation of the 11-dimensional
supermembrane world volume theory. We explicitly construct vertex operators for the
massless states and study interactions of supermembranes. 
The open supermembrane and its vertex operators are discussed.   
We show how our results have direct applications to Matrix theory
by appropriate regularisation of the supermembrane. 

\newpage
\section{Introduction}
As we review the progress over the last few decades we find that there are 
unexplored avenues in the 11 dimensional supermembrane
theory which can lead to a better understanding of quantum
M-theory and hence non-perturbative string/quantum gravity.
The supermembrane is a 2+1 dimensional object moving in
11 dimensional space, with a world volume theory \cite{bst} which
when quantised will give us a glimpse of the fundamental 
degrees of freedom of M-theory. Attempts to quantise the world volume theory
in analogy to 10 dimensional world sheet string theory
have revealed many interesting features which distinguish
it from the 10 dimensional string theory, and also make
it difficult to solve. The main source of difficulties lies in the fact 
that, unlike the string, the 2+1 dimensional world volume theory 
is interacting. Moreover it does not have conformal invariance.
Certain features of the supermembrane are understood, namely, it 
can be regularized
 to yield a supersymmetric 
Matrix theory \cite{dwhn1}, and it's 
spectrum is continuous \cite{dwhn2,sm}. These
lead to a multiparticle interpretation of the spectrum \cite{bfss}, 
and the need
for a second quantised description of the membrane. We still do not have a
complete understanding of this, nor is the existence of a
normalisable ground state confirmed. However, there is evidence
from Matrix theory that such a state indeed exists and it contains
massless states corresponding to the massless multiplet of 11 dimensional supergravity\cite{gs}. 

Given this situation, we can look for further quantities to
describe the supermembrane which will hopefully shed some more 
light on the underlying quantum theory.
The interactions of the massless sector is one interesting
avenue to explore. There are computations of $d=11$ scattering
amplitudes for the massless sector \cite{ggk,pw,gg} using the 
quantised superparticle.
However the calculated amplitudes diverge, and the coefficients
can be fixed only by duality with 10 dimensional string theory. 
As we know, in
the case of the superstring \cite{ssv} as well as the 
superparticle \cite{ggk,gg}
scattering amplitudes are evaluated by determining the 
vertex operators and inserting them into path-integral amplitudes. 
The vertex operators for the supermembrane derived in \cite{dnp}, precisely seek to
achieve the same for a supermembrane scattering amplitude \cite{pl,bnpw}. The operators are determined uniquely and provide the first step towards understanding
massless interactions. 
Both the superstring and the superparticle vertex operators are contained in the supermembrane
vertex operators. By reducing the two spatial dimensions of the supermembrane world volume to zero,
the superparticle is obtained. (This we call the particle limit). On the other hand, wrapping
one of the directions of the membrane along a compact direction gives us 
a superstring in $d=10$.
This `double dimensional reduction (DDR)' of the $d=11$ 
supermembrane vertex operators gives us
all the superstring vertex operators. Also, by virtue of the 
fact that Matrix theory is recovered after appropriate regularisation, 
these vertex operators give linearised Matrix action in weak
$d=11$ supergravity background.  

\begin{figure}
\begin{center}
\epsfxsize=12cm
\epsfbox{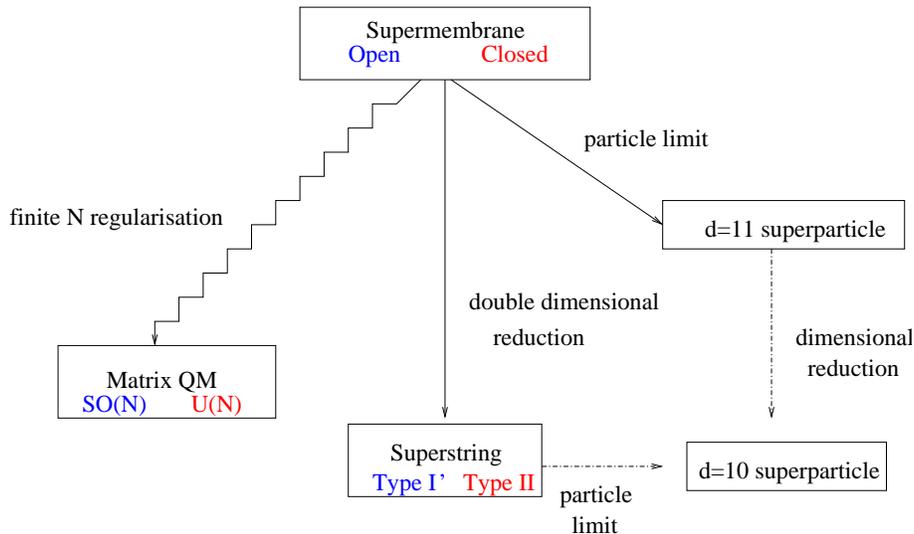}
\end{center}
\caption{Various limits of the supermembrane model}
\end{figure}


 We also discuss the vertex operators for the open membrane as it is closely related
to heterotic string theory \cite{hw} and in recent times 
has assumed importance due to its relation to non-commutative
theories in presence of a background constant three form
gauge field strength \cite{opm}. 
 Many other topics regarding the quantum supermembrane
have been discussed in detail in previous reviews
\cite{revs,nh}.

The first section is an introduction to
supermembrane basics, mainly to fix notations.
We also review the relation of supermembrane
to Matrix theory, and then the multiparticle
interpretation of the supermembrane spectrum. 
In the second section we give the vertex operators for the
massless sector of the theory.  
The third section deals with scattering 
amplitude calculations with the help of vertex operators.

\section{Supermembranes}
In analogy with the particle and the string, the action for 
the membrane is given by the 2+1 dimensional
world volume swept out in the target space. The supermembrane is a 
extension obtained by adding fermionic target space `coordinates'. 
A supersymmetric action involving the bosonic and the fermionic
coordinates can be written down consistently only in special dimensions, of which d=11 is the maximal value. \footnote{There
are other formulations of the action using twistors \cite{tw}.} 
This action in the background of $d=11$ supergravity takes the form


\begin{eqnarray}
S= T \int d^3\xi \left[-\frac12\sqrt{-g} g^{ij}\Pi^a_i\Pi^b_j\eta_{ab} + \frac12\sqrt{-g} -\epsilon^{ijk}\Pi^A_i\Pi^B_j\Pi^C_k B_{CBA}\right].
\label{cuact}
\end{eqnarray}

 Superspace notation has been used where space comprises of both the 
bosonic and fermionic coordinates. Explicitly 
\begin{eqnarray}
\Pi_i^A &= &\partial_i Z^M E^A_M \quad  A=(a,\alpha); \ \ \ 
a=0,\ldots,10  \quad \alpha= 1,\ldots, 32 \nn \\
Z^M&=&(X^\mu.\theta^{\dot\alpha})\quad \,  M=(\mu, \dot{\alpha}) \ \ \ 
\mu=0,\ldots,10 \quad \dot{\alpha}=1,\ldots, 32  \nn 
\label{ac}
\end{eqnarray}
where $Z^M$ are the coordinates of the membrane in the target 
superspace, $E^A_M$ is the super-elfbein, and $\xi^i$,
($i=0,1,2$) denote the 2+1 dimensional world volume coordinates. 
Tangent space indices are denoted by $A$ and world indices by $M$.
$g_{ij}$ is the world volume metric ($g={\rm det}\, g_{ij}$) which 
is considered independent here, and $B_{CBA}$, the three form of $d=11$ 
superspace. $T$ is the tension of the supermembrane, and it is the sole
length scale in the theory. Unlike the superstring coupling constant 
$g_s$ there is no analogous supermembrane coupling constant. 
This precludes a perturbative expansion of supermembrane
interactions making the supermembrane theory non-perturbative theory from the very outset. We shall set $T=1$ for convenience henceforth.

The action has redundant degrees of freedom due to
the presence of the local reparametrisation
freedom of the world volume and fermionic $\kappa$ symmetry of the 
action \cite{bst}. The $\kappa$ symmetry of the action requires
one crucial fact : the component fields of
the graviton  $h_{\mu\nu}$, gravitino $\Psi_{\mu}$ and the three
form $C_{\mu\nu\rho}$ obey $d=11$ supergravity equations of motion
\cite{csj}. This is similar to the GS formulation of the superstring, 
but quite different from the NSR superstring, where one has to resort
to $\beta$ function evaluation to show that the background fields obey
the supergravity equations of motion.  
However, one 
aspect of the action (\ref{cuact}) is that though it gives the coupling 
of the supermembrane to superspace background fields, it is very difficult 
to extract the coupling in terms of the {\it component} fields 
(i.e. $h_{\mu\n}, \Psi_{\m}, C_{\m \n\r}$). The method of gauge completion 
used in \cite{dpp}, has succeeded in determining the couplings only to 
order $\theta^3$ of the target space
fermions. There are other methods of evaluating the action in components 
as in \cite{kg}, but, as we show in the next section
to linear order in the fields, the couplings can be determined exactly, 
through the evaluation of vertex operators.

To begin with, one restricts oneself to quantising the 
supermembrane propagating in {\it flat} target space. The coordinates
in $d=11$ flat space are $Z^M= (X^{\mu},\theta^{\alpha})$, $\theta$ 
is a Majorana spinor. The superelfbein can be determined \cite{bst}, 
and we have 
\begin{eqnarray}
\Pi_i^{\mu} & = &\partial_i X^{\mu} - i  \bar\theta\Gamma^{\mu}\partial_i\theta,  \ \ \Pi^{\alpha}_i=\partial_i\theta^{\alpha}, \nn\\ B_{\mu\nu\rho} & = & 0, B_{\mu\nu\alpha}= -\frac{i }{6}\left(\Gamma_{\mu\nu}\theta\right)_{\alpha} , B_{\mu\alpha\beta} = -\frac{1}{6}\left(\Gamma_{\mu\nu}\theta\right)_{(\a }\left(\G^{\n}\th\right)_{\b)}, \nn \\ B_{\a\b\g} & = &-\frac{i }{6}\left(\G_{\m\n}\th\right)_{(\a }\left(\G^{\m}\th\right)_{\b}\left(\G^{\n}\th\right)_{\g)}.
\end{eqnarray}
 In addition, we use the equation of motion for the world volume 
metric $g_{ij}=\eta_{ab} \Pi^{a}_i\Pi^b_j$. Plugging this into
(\ref{cuact}) gives the supermembrane in flat space 
\begin{eqnarray}
S&=&-\int d^3\xi \le\{\sqrt{- g(X,\theta)} - \epsilon^{ijk}\bar{\theta}\Gamma_{\mu \nu} \partial_i\theta \nn\re. \\ &&\le.\left[\frac12\partial_jX^{\mu}\left(\partial_k X^{\nu} + \bar{\theta}\Gamma^{\nu}\partial_k\theta\right) + \frac16\bar{\theta}\Gamma^{\mu}\partial_j\theta \bar\theta\Gamma^{\nu}\partial_k\theta\right]\re\}.
\end{eqnarray}
The reparametrisation of the world volume can be used
to fix the bosonic degrees of freedom to eight. (E.g.\ 
one can go to static gauge
by identifying
$X^{0,1,2}=\xi^{0,1,2}$ leaving 8 transverse degrees). The $\kappa$
symmetry of the action is then used to reduce the number of fermionic degrees of freedom. 
 The most efficient way of dealing with the above action and its symmetries however
is to go to the lightcone gauge \cite{bst,dwhn1}. The light cone directions $X^{\pm} =\frac1{\sqrt2}(X^0 \pm X^{10}) $
are singled out, and $X^+$ is identified with the time direction of the world volume, which we denote by $\tau$
here (hereafter, we denote the spatial directions $\xi^{1,2}\equiv\sigma^{1,2}$). The $\kappa$ symmetry
also allows to set 1/2 of the fermionic degrees of freedom to zero. The gauge conditions thus read:
\begin{eqnarray}
X^{+}& = &X_0^{+} + \tau\\
\Gamma^{+}\theta &=&0,
\label{ga}
\end{eqnarray}
where $\Gamma^{\pm}=\frac{1}{\sqrt 2}\left(\Gamma^0\pm\G^{10}\right)$. 
One is left with 9 transverse degrees of freedom and the
fermionic coordinate $\theta$ is reduced to having 16 degrees
of freedom in this gauge. In addition one has to solve for
$X^{-}$ which gives an extra constraint reducing the total
number of bosonic degrees of freedom to 8. The details of this
derivation can be found in \cite{bst,dwhn1}.

The Hamiltonian density derived from the gauge fixed lagrangian 
has the following form ($P^+= \d{\cal L}/\d\partial_{\tau} X^-; \vec P= \d {\cal L}/\d\partial_{\tau}\vec X$):
\begin{equation}
{\cal H}= \frac{1}{2 P^{+}}\left(\vec P^2 + \bar g\right) - \epsilon^{rs}\partial_rX^a\theta\gamma^a\partial_s\theta
\label{hamil}
\end{equation}
where $P^{+}$ can fixed as a constant of motion, $\vec P= \partial_{\tau} \vec X$, 
$\bar g= (\epsilon^{rs}\partial_{r}X\partial_{s}X)^2$ 
and $a,b,..$ denote the transverse directions.
We also use $SO(9)$ $\g$ matrices henceforth. 
It is noteworthy that the Hamiltonian does not depend on the centre of mass coordinates $\vec X_0$ and $\theta_0$.
The supermembrane mass formula 
\be
{\cal M}^2 = 2P^+P^- -\vec P_0^2 
\label{mass}
\ee
is also independent of the centre of mass coordinates.
The Hamiltonian is accompanied by a constraint (obtained by taking curl of the equation $\partial_r X^-= \partial_{\t}X^a\partial_rX_a - \i \th\partial_r\th$
\begin{equation}
\phi= \epsilon^{rs}\left(\partial_r{\vec P}\partial_s{\vec X} - \i \partial_r \theta\partial_s\theta\right)\approx 0
\label{const}
\end{equation}
The potential in the Hamiltonian
\begin{equation}
\bar g = {\rm det} \, (\partial_r {\vec X}\partial_s{\vec X}) 
= (\epsilon^{rs}\partial_r X\partial_s X)^2 
\end{equation}
vanishes for configurations where one of the directions is independent of one of the membrane world volume
coordinates. In other words there are valleys in the potential signalling degenerate surfaces. The membrane
surface can be made arbitrarily narrrow in one direction, and hence a single membrane can be deformed into
arbitrary number of membranes without any cost of energy. 
This is interpreted as the non-existence of
a definite membrane number, and no definite membrane topology. A path integral approach to membrane
quantisation like in string theory ({\it \`a la} Polyakov),  where there is a sum over different topologies for the
world sheet, cannot be achieved. 

The light cone Hamiltonian is still invariant under
a class of diffeomorphisms, called area-preserving diffeomorphisms (APD), 
acting as \cite{GH,dwhn1}:
\begin{equation}
\delta X^\mu= -\xi^r\partial_r X^\mu\qquad 
\delta \theta= -\xi^r\partial_r \theta
\end{equation}
with $\xi^r=\epsilon^{rs}\partial_s\xi$, such that $\partial_r\xi^r=0$ 
($\xi$ is a scalar parameter) , and hence area of the 
two surface is preserved under this transformation.
The area preserving diffeomorphisms as defined above can be 
attributed with the following Lie bracket
structure:
\begin{equation}
\delta A= \{\xi,A\} = \epsilon^{rs}\partial_r\xi\partial_s A.
\end{equation}
Thus $\{A,B\}=\epsilon^{rs}\partial_rA\partial_s B$ defines a Lie bracket 
for any two functions $A,B$. It shares all the requisite properties
of Lie bracket, namely antisymmetry, associativity and satisfies the
Jacobi identity. The Hamiltonian density can thus be rewritten as: 
\begin{equation}
{\cal H}= \left(\vec P\cdot \vec P -\frac12\left\{X^a,X^b\right\}^2 + \theta\gamma^a\left\{X^a,\theta\right\}\right)
\label{hapd}
\end{equation}

In fact, with the above knowledge, one can start with a lagrangian 
invariant under APD, to yield (\ref{hapd}) as the Hamiltonian. 
The APD invariant Lagrangian can be written in a compact form 
by introducing an auxiliary field
$\omega$, which transforms as a gauge field under APD transformations:
\begin{equation}
\delta\omega= \partial_{\tau}\omega + \left\{\xi,\omega\right\}
\end{equation}

By defining a covariant derivative, 
$$ DX^a = \partial_{\tau} X^a -\left\{\omega,X^a\right\}, $$
the lagrangian becomes \cite{dwhn1}:
\begin{equation}
{\cal L}= \frac12\left(D X^a\right)^2 -\frac14\left\{X^a,X^b\right\}^2 -i  \theta D \theta -i  \theta \gamma^a\left\{X^a,\theta\right\}
\end{equation}
The lagrangian is also invariant under the target space supersymmetries, 16 of which are linearly realised with parameter $\eta$, and due
to the gauge fixing, another 16 of which are non-linearly realised with parameter $\epsilon$. 
They are of the form:
\begin{eqnarray}
\delta X^a &= &-2\epsilon \gamma^a \theta \;\;\;\; \delta\omega = -2\epsilon \theta  \nonumber\\
\delta \theta &= &i  D X^a \gamma_a \epsilon -\frac{i }2 \left\{X^a, X^b\right\} \gamma_{ab} \epsilon+ \eta
\label{sutr}
\end{eqnarray}

Note that the action is invariant under the above transformations
up to total derivatives, and for closed supermembranes the total 
derivatives do not make any contribution (The fields are 
assumed to vanish at $\tau=\pm\infty$). However, for open supermembranes, 
where there are boundaries at the end of spatial 
directions of the world volume,
 Dirichlet or Neumann boundary conditions have to be imposed to ensure
that the boundary terms vanish. One finds that to ensure 
invariance under supersymmetry, it is
necessary that the supermembrane ends only on 1, 5 or 9-dimensional 
hypersurfaces. This set of conditions have been derived 
earlier in \cite{eza,dwpp} and by demanding invariance under
$\kappa$ symmetry of the covariant action \cite{ce}.

To impose the boundary conditions on the ends of the supermembrane, we define the normal and tangential derivatives on the
boundary:
\begin{eqnarray}
\partial_n X^{\mu} &\equiv &n^r\partial_r X^{\mu} \\
\partial_t X^{\mu} &\equiv &\epsilon^{rs}n_r \partial_s X^{\mu}
\label{nn}
\end{eqnarray} 
where $n^r$ is the unit normal on the boundary and $\epsilon^{rs}n_r$
the unit tangential vector to the boundary.
For the supermembrane ending on a $p$ dimensional hypersurface 
the following boundary conditions are required:
\begin{eqnarray}
\partial_n X^M &= & 0 \quad \mbox{for} \; M=2,\ldots,p \quad \mbox{(Neumann)}\\
\partial_tX^m & = & 0 \quad \mbox{for}\; m=p+1,..,10 \quad \mbox{(Dirichlet)}
\label{bndry}
\end{eqnarray}

To check for invariance under supersymmetry transformations (\ref{sutr}), we vary the action and find the following boundary terms:
\begin{eqnarray}
\delta {S}&=& -\int d{\tau} \int_{\partial G} d\sigma 
 \eta \gamma^a\theta \partial_t X^{a} + \nonumber \\
&& + \int d\tau \int_{\partial G} d\sigma \epsilon \gamma^{d}
\left(\gamma\cdot DX - \frac12 \gamma^{ab}
\left\{X^a,X^b\right\}\right)\theta \partial_t X^{d}.
\end{eqnarray}
On imposing the (\ref{bndry}), we find
to get the terms to vanish along the Neumann directions, additional conditions on $\theta$ must be imposed.
These translate as
\be
\eta\gamma^M\theta=\e \g^M\g^N\th= \e\g^M\g^m\g^N\th=0. 
\label{opn}
\ee
Defining $P_{\pm}= 1/2(1 \pm \gamma^{p+1}\ldots\gamma^{10})$, which act as projection operators for
dimensions $p=1,2,5,6,9$, we find that the following conditions:
\begin{equation}
P_{-}\theta=0\ \ \ 
P_{+}\epsilon=0 \ \ \
P_{-}\eta=0
\label{bnd}
\end{equation}
are required so that (\ref{opn}) is obeyed, which restrict $p=1,5,9$. It is interesting to see that
(\ref{bnd}) results in the fermionic degrees of freedom being reduced to 8 on the boundary. For
$p=9$, $P_{-}$ coincides with the chiral operator for the boundary theory. This is the first sign
that the boundary theory, which is essentially a string theory induced by the membrane has a
heterotic structure. In fact, by looking at the equations of 
motion obeyed by the membrane on the boundary, we find 
that they are `free' equation of motion for a string. For simplicity,
we discuss the $p=9$ case, and its relation to heterotic Matrix theory \cite{het}. The $p=5$
case also has many interesting applications \cite{five}, 
especially in the light of non-commutative
open membrane theories proposed and discussed in \cite{opm}. The $p=1$ case is yet to be investigated.
 
Once we have ensured that the boundary terms vanish, the bulk equation of motion for
the fields are:
\bea
D^2 X^a -\le\{\le\{X^a,X^b\re\},X^b\re\} - \i\le\{\th,\g^a\th\re\} &= &0 \label{xeq}\\
D\th + \le\{\g\cdot X,\th\re\} &= & 0 \label{theq}\\
\le\{D X^a, X^a\re\} - \i \le\{\th,\th\re\} & = &0
\eea

The last of these, the equation of motion for the auxiliary gauge 
field $\omega$ is same as the constraint (\ref{const}). What is 
interesting is how these equations reduce on the boundary. Taking
$a=m$ in (\ref{xeq}) on the boundary, the normal derivative becomes:
\be
\partial_n X^m= {\mbox constant}
\ee
Using this, we define $\bar{\g}= \sum_{p+1}^{10} \partial_n X^m\g_m=$ 
constant matrix. The equations obtained by putting $a= M$ in 
(\ref{xeq}, \ref{theq}) reduce to the following linear wave equations
on the boundary:
\bea
(\partial_{\t}^2 - \partial_t^2) X^M &= & 0 \\
(\partial_{\t} - \bar\g \partial_t)\th &=&0
\eea
(recall that $\partial_t$ is the tangential derivative along the spatial
boundary of the membrane). In this way we see that the supermembrane 
equations of motion with the above boundary conditions induce a 
superstring theory on its boundary. The restrictions on the value of $p$
can then be easily understood, because only for these values is it possible
to match bosonic and fermionic degrees of freedom on the boundary.
In particular, for $p=9$, we obtain the world sheet equations 
for the heterotic string, and $\bar \g=\g^{11}$ becomes a chirality 
matrix on the (9+1)-dimensional brane.
By (\ref{bnd}), only one of the chiralities of $\th$ survives on the boundary,
leaving an appropriate equation of motion for the chiral coordinate. 
We had to put $\partial_n\th=0$ to obtain the free heterotic string equations 
of motion. 

After we have obtained the gauge fixed lagrangian, with the appropriate 
equation of motion for the world volume fields, what remains is the 
quantisation of the theory. Some attempts in this direction have been 
reviewed in \cite{nh}. Here we confine ourselves to brief comments 
and show how Matrix regularisation of the membrane leads to a
interpretation of the spectrum. We describe the latter first.
\subsection{In the light of Matrix theory}

We now concentrate on the relation of the APD diffeomorphism transformations to SU(N) (or SO(N) for the open membrane) 
gauge transformations
under suitable regularisations \cite{dwhn1}.
It was first discussed in \cite{dwhn1}, and led to the supermembranes' relation to
Matrix theory, and M theory \cite{bfss}. In the case of open membranes, it leads to the
relation to heterotic Matrix theory for membranes ending on 9-branes \cite{hw}, and
to other exotic theories \cite{opm}.

Since this discussion appears in previous reviews \cite{nh}, we just
briefly give the relation here, with an emphasis on the regularisation of the open membrane. 
As stated earlier, the APD bracket has a Lie bracket structure. It is also easy to check that
the commutator of two APDs leads to a third APD:
\begin{equation}
\{\xi_2,\xi_1\}= \xi_3
\end{equation} 
Given a basis of orthogonal functions $\{Y^A\}$ on the spatial manifold, we 
can expand the coordinates as $X^a(\sigma) = X^a_0 + \sum_{A}X^{aA} 
Y_A(\sigma)$. The Lie bracket then assumes the following form:
\begin{equation}
\{Y_A,Y_B\}= g_{ABC} Y^C \ \ \  Y^A=\eta^{AB} Y_{B} \ \ \eta^{AB}\eta_{BC}= \delta^A_C
\label{lie}
\end{equation}
where 
\begin{equation}
g_{ABC}= \int d^2\sigma \epsilon^{rs}Y_A\partial_rY_B\partial_sY_C
\qquad \eta_{AB}=\int d^2\sigma\, Y_A Y_B 
\end{equation}

Given the above, the basis which is infinite dimensional for a continuum
manifold, can be restricted to some finite $A=1,..,\Lambda$ such that
\begin{equation}
\mbox{lim}_{\Lambda\rightarrow\infty} f_\Lambda^{ABC}= g^{ABC}
\end{equation}
with $f_\Lambda^{ABC}$ the structure constant of some finite 
dimensional group labeled by $\Lambda$.
For closed membranes of arbitrary topology \cite{dwhn1,mat} this finite
dimensional group is SU(N), and the Hamiltonian of the regulated
membrane turns out to precisely coincide with the Hamiltonian
of dimensionally reduced SU(N) super-Yang Mill's theory. This can easily be seen by substituting
the regularized coordinates in (\ref{hapd}), and replacing the APDs by
appropriate commutators. The non-zero $X_A$'s transform in the adjoint
representation of the SU(N) group.
\begin{equation}
H=Tr\left(\frac12 + \frac14 \left[{\bf X}^a,{\bf X}^b\right]^2 + \left[{\bf X}^a,\bar {\Theta}\right]\gamma^a{\Theta}\right)
\end{equation}
In the above $({\bf X}, \Th)$ denote matrices.
The same Hamiltonian was used in \cite{bfss} to describe $N$ $D0$-branes 
in the infinite momentum frame, with their momenta along the 11th direction
of $d=11$ space. Hence supermembranes through Matrix theory are 
intimately related to $D$-branes and M-theory.

For open supermembranes, the matrix regularisation yields different 
finite dimensional groups, dependent on the topology of the 
continuum membrane:  for the disc $D^2$, the cylinder, and the 
M\"obius strip, we get the groups $SO(N)$, whereas for the 
projective plane we get $USp(2N)$ \cite{rey}. 
Moreover, depending whether the $X^a$ are Neumann or Dirichlet,
they either transform in the symmetric or adjoint representation
of the gauge group. We shall illustrate here with the example of the
disc, the regularisation of the open membrane.
Consider $Y_A= Y_{lm}(\theta,\phi)$, or spherical harmonics, with 
$m\leq |l|$. The restriction that $l\leq N-1$ for the spherical
membrane leads to a basis with $\sum_{l}^{N-1} (2l + 1)= N^2-1$ 
independent components, and the ${\bf X}$ transform in the
adjoint of the SU(N) gauge group. However, when the membrane is 
stuck on a hypersurface,
it essentially corresponds to a disc topology, with 
boundary conditions imposed on the $X^m$ ($m=p+1,..10$), 
this would translate as $ K^-_{lm}= Y_{lm} - (-)^{(l+m)}Y_{lm}$
for $l+m$ odd being the correct basis. This gives $\sum_l^{N-1}l = N(N-1)/2$
as the number of generators of the finite group. This as we know
is the number of generators of SO(N).  For the Neumann directions (\ref{nn}),
 we get these directions to transform {\it symmetric} tensor
representations of SO(N) \cite{dwpp,eza}. In fact, the regularised
action has the following form (in 11 dimensions):
\bea
S &=& \int Tr \left( D{\bf X}^2 + DA_{10} + [{\bf A_{10}}, {\bf X}^i]^2 + [{\bf X}^i,{\bf X}^j]^2 \re. \nn\\ &&\le. -\i \Th^+ D\Th ^+
-\i \Th^-D\Th^- + 2\i \Th^+\G^i[{\bf X}^i,\Th^-]\right)
\eea
Where we have distinguished $X_{10}=A_{10}$ as it transforms 
in the adjoint of the SO(N) group. Also the fermions are broken up as $\Theta^+, \Theta^-$ which transfrom in the adjoint and symmetric representaion of the gauge group. The ${\bf X^i}$ transform 
in the symmetric traceless representation of SO(N). 
The above matrix regularisation is the heterotic
Matrix theory \cite{het}, but without the twisted sector fields
expected to yield the additional $E_8\otimes E_8$ degrees of 
freedom \cite{hw}. These twisted fields appear at the boundaries 
of the membranes, and live only on the 9-branes.
A proper membrane origin of these fields is yet to be determined.
For the case of the membrane ending on a five brane, there are many 
interesting possibilities \cite{five}, but much remains to be done.

The resultant theory is yet to be quantised fully. However, from 
the nature of the
Hamiltonian, it can be seen that the supermembrane spectrum is continuous
and there is no mass gap. This points towards a multiparticle interpretation
of the spectrum, and hence a second quantised picture of the supermembrane.

\subsection{Approaching quantisation}
The matrix regularisation of the supermembrane proves very useful,
as one can use matrix quantum mechanics and interpret the
$N\rightarrow \infty$ as the quantum supermembrane.
However, the
$N\rightarrow \infty$ limit is very subtle (membranes
with different topology and hence different APD are approximated
by  the
same SU(N) regularisation). But the multiparticle interpretation of the supermembrane spectrum
comes entirely from its relation to Matrix theory. It was shown in \cite{dwhn2}
that the supermembrane spectrum is continuous and there
is no-mass gap. 
The continuous spectrum of the supermembrane was interpreted
initially as a signature of instability. 
However, now due to its relation to matrix theory,
this is attributed to the presence of multi-particle
states. In the original conjecture of \cite{bfss} , the diagonal 
elements of $SU(N)$ matrices are positions of $D0$ branes, 
and in the case of block diagonal
matrices, each block corresponds to $N_i$ coincident $D0$-branes,
where $N_i$ is the dimension of the $i$th block.
Each of these can be thought as separate entities, and in the
infinite $N$ limit, as separate membranes linked by thin tubes.  
This multiparticle interpretation of the supermembrane spectrum
implies, that we should essentially treat the supermembrane
world volume as a second quantised theory.

\begin{figure}[t]
\begin{center}
\epsfxsize=13cm
\epsfbox{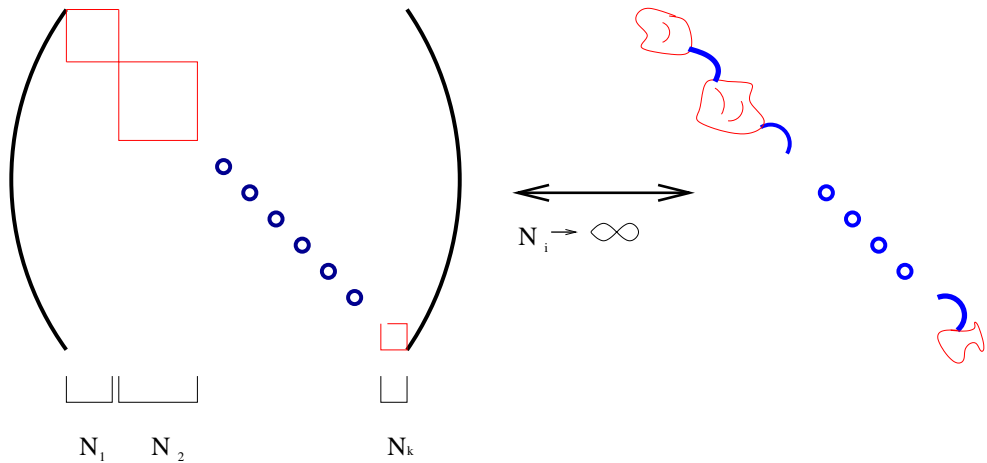}
\end{center}
\caption{Matrix-membrane correspondence}
\end{figure}

However, one crucial question still remains: can one find
a normalisable ground state for the theory? And if so,
does the spectrum have massless states? By just looking
at the zero mode sector of the theory, one can build
states which transform as ${\bf 44} \oplus {\bf 88}$ of SO(9)  
in the bosonic sector and $\bf{128}$ of SO(9) in the fermionic sector. 
Since the Hamiltonian and hence mass (\ref{hamil}) does not depend 
on the zero modes, this should
give the massless sector of the theory, provided the groundstate
corresponding to the non-zero modes transforms as a SO(9)
singlet. Attempts to prove this have not seen much success up to now.
However, in Matrix theory, considerable progress has been made in 
efforts to prove the existence of normalisable ground states
\cite{gs} in the case of SU(2) or SU(3). 
Further information  can be found in the review \cite{nh}
and we refrain from giving the details here. In the next section,
we discuss vertex operators for the supermembrane, which is one
of the directions which can shed more light on this side of M-theory.

\section{Vertex Operators}
Though it still remains to be proven that the
supermembrane has massless states, there is sufficient evidence 
that they exist, and as discussed in previous 
sections, should belong to the $d=11$ supergravity
multiplet, obeying appropriate onshell conditions. This then leads to the
question whether there exists vertex operators for the supermembrane
in analogy with the superstring vertex operators. The superstring vertex
operators prove to be extremely useful in the calculation of
string interactions due to a number of reasons.
The superstring is a first quantised theory, and hence vertex operators
which are local operators act as `creation' operators. 
Moreover, the conformal symmetry of the string allows the asymptotic
states of the spectrum to be mapped to local operators at finite points on the world sheet.
The vertex operators can then be inserted in the path integral, and 
of scattering amplitudes evaluated using them.
Further, conformal invariance also restricts the nature of the vertex operators
which can are uniquely determined.

Vertex operators by definition are of the following form:
\begin{equation}  
V_{h}= \int d^3\xi\,  h\cdot O_h\left[X,\theta\right]e^{i  k \cdot x}
\label{ver}
\end{equation}
where $h$ denotes the polarisation of the state, and $O$ the local operator corresponding to that state which has a momentum $\vec k$. Given the fact that supermembrane world volume theory does not have conformal invariance we have to look elsewhere to determine $O$. 
However, the requirement that the vertex operators transform into one another
under under supersymmetry transformation competely determines their
structure.
In addition the technique of double dimensional reduction to 
give the superstring
in lower dimensions gives us an additional check for the supermembrane
vertex operators. The superparticle vertex operators for $d=11$ were
determined in \cite{ggk}, and also serve as a useful guide in our 
calculations. 

Thus for the supermembrane the vertex operator for the graviton should be of the form (\ref{ver})
\begin{equation}  
V_{h}= h_{a b}\int d\tau d^2 \sigma O^{ab}_h\left[X^a(\tau,\sigma_i),\theta(\tau,\sigma_i)\right]e^{i  k \cdot x}
\end{equation}
$h_{ ab}$ denotes the
polarisation tensor of the graviton, and $O^{ab}$ is a local operator
of the supermembrane light cone coordinates. $\vec k$ denotes the momentum
of the graviton, and this operator in the superstring case creates a
graviton state when acting on the string ground state.  Using the fact that
under supersymmetry transformations
\begin{equation}
\delta V_h= V_{\delta \psi(h)}
\label{gs}
\end{equation}
where $\delta \psi_{h}$ is the transformed gravitino polarisation vector,
we can solve for $V_h$. And $V_{\psi}$, $V_C$ which have the following
transformations under supersymmetry:
\begin{equation}
\delta V_{C}= V_{\delta \psi(h)} \ \ \ \delta V_{\psi}= V_{\delta h} + V_{\delta C}
\label{gcs}
\end{equation}

Since we work in light cone gauge the polarisations are separated as transverse
(with indices a,b) and longitudinal (with indices $+,-$). Further the gravitino is split up into two SO(9) spinors $\psi_a,\tilde\psi_b$.
To solve for the above vertex operators, we have to resort to $d=11$ 
onshell conditions. 
Note that these conditions are ensured by the $\kappa$ invariance of the
supermembrane action \cite{bst}.
These conditions are given by
\bea
&k^a h_{ab} = h_{aa} = 0 = k^a C_{abc}&\nn \\
&\g^a \,\tilde\psi_a = \g^b k_b\, \tilde\psi_a = k^a\,\tilde\psi_a
= 0 =k^a\,\psi_a &\nn\\
&\g^a\, \psi_a=\tilde\psi_+ \qquad k^b \g_b\, \psi_a= k^-\, \tilde\psi_a
\label{sugraconst}
\eea
Because the polarisations are given in light cone gauge, the 
supersymmetry transformations (\ref{gs},\ref{gcs})
are separated into 16 linearly realised ($\eta$) and 
16 non-linearly realised ($\epsilon$). However for the sake
of simplicity, we shall here record only on the linear supersymmetry
transformations listed below
\bea
&\delta h_{ab}  = -\tilde\psi_{(a}\g_{b)}\eta  \qquad
\delta h_{a+} = -\ft 1 {\sqrt 2}\, \psi_a\eta \label{dhab}&\\
&\delta C_{abc} = \ft 32 \tilde\psi_{[a}\g_{bc]}\eta \qquad
\delta C_{ab+} = \sqrt 2\, \psi_{[a}\g_{b]}\eta \label{dCabc}&\\
&\delta \psi_{a} = k_{b}\, h_{ca}\, \g^{bc}\eta + \ft {1}{72}
(\g_{a}{}^{bcde}\, F_{bcde}- 8 \g^{bcd}\,F_{abcd})\, \eta&\\ 
&\delta \tilde\psi_+ =-\ft {\sqrt{2}}{72} \g^{abcd}\eta\, F_{abcd} \label{gp}
\delta \psi_+ =
\delta \tilde\psi_{a} =0=\delta h_{++}&\nn
\eea

We now state the vertex operators, and shall explicitly check for the
linear supersymmtery of the graviton vertex operator ($ R^{abc}=\frac1{12}\th \g^{abc}\th,\ \ R^{ab}= \frac1{4}\th\g^{ab}\th$).
\bea
V_h&=& h_{ab}\, \le[ DX^a\, DX^b - \{X^a,X^c\}\,\{X^b,X^c\} - \i
\theta\gamma^a\,\{X^b,\theta\} \re. \nn\\
&& \le.-2 DX^a\, R^{bc}\, k_c - 6 \{X^a,X^c\}\, R^{bcd}\, k_d
+ 2 R^{ac}\,R^{bd}\, k_c\, k_d \re]\, e^{-\i k\cdot X} \label{Vh}\\
V_{h_+} &=& -2 h_{a+} \, (DX^a - R^{ab} k_b) e^{-\i k\cdot X} \label{Vh+}\\
V_{h_{++}} &=&  h_{++}\, e^{-\i k\cdot X}
\label{Vh++}
\eea
\bea
V_C &=& -C_{abc}\, DX^a\, \XX{b}{c}\, e^{-\i k\cdot X} + F_{abcd}\le[
(DX^a-\ft 23 R^{ae}\, k_e)\, R^{bcd}\re. \nn\\
&&\le. -\ft 12 \XX{a}{b}\, R^{cd} -\ft 1 {96}\, \XX ef\, \theta \gamma^{abcdef}
\theta \re]\, e^{-\i k\cdot X} \label{VC}\\
V_{C_+} &=& C_{ab+} (\XX a b +3 R^{abc}\, k_c) e^{-\i k\cdot X}
\label{VC+}
\eea
\bea
V_\Psi &=&
\psi_a\, \le[ \, \le(DX^a-2 R^{ab}\, k_b +\gamma_c\,
\XX ca\, \re) \theta \re]\, e^{-\i k\cdot X}\nn\\
&&+ \tilde\psi_a\, \le[ \gamma\cdot DX\,  \le (DX^a- 2R^{ab}\, k_b
+\gamma_{c} \XX ca \re)\,\theta\re. \nn\\
&&+\ft 12 \gamma_{bc}\, \XX b c \, ( DX^a- \XX a d \, \gamma^d\, )\theta
+ 8  \gamma_b\theta\, \XX b c \, R^{cad}\, k_d \nn\\
&&+ \ft 5 3 \gamma_{bc}\theta\, \XX b c R^{ad}\,k_d
+ \ft 4 3 \gamma_{bc}\theta\, \le( \XX a b\, R^{cd} + \XX c d R^{ab}
\re) k_d \nn\\
&&
\le.+\ft 2 3 \i \,\le ( \gamma_b\theta\, \{X^a,\theta\}\gamma^b\theta
-\theta\, \{X^a,\theta\}\theta \,\re )
+ \ft 8 9 \gamma^b\theta\, R^{ac}\, R^{bd}\, k_c\, k_d
\,\re] \, e^{-\i k\cdot X} \label{VPsi}\\
V_{\Psi_+} &=& -\left[\psi_+  \th + \tilde\psi_+ \left( \g^a DX^a +
             \ft12 \g^{ab} \{ X^a,X^b\} \right) \th \right] e^{-\i k\cdot X}
\label{VPsi-}
\eea

To see that the linear supersymmetry is realised, 
 we implement the following transformation in the graviton vertex (\ref{Vh}):
$\delta X^a=\delta\omega=0, \delta \theta=\eta$. In otherwords only the
terms proportional to $\theta$ shall contribute to the variation, which
is written thus:
\bea
\delta V_h & = &k_b h_{ca}\eta\gamma^{bc}\left[DX^a - 2 R^{ad}k_d -\gamma^d\XX ad \theta\right] e^{-i  k\cdot x} \nn \\ &&- h_{ ab}\left[\{X^a,k\cdot X\}\eta \gamma^b\theta + i  \eta\gamma^a\{X^b,\theta\}\right]e^{i  k \cdot X}
\eea

The terms in the first line can be grouped together to yield the gravitino
polarisation as per (\ref{gp}). The terms in the second line have to vanish
clearly, and by a partial integration cancel each other. Note that, in the
case of the open membrane we have to be careful in order to ensure the
vanishing of the above. In fact, we find that the additional condition 
of $\partial_t \omega \Big|_{\partial G}=0 \, , \, h_{mM}\Big|_{\partial G} =0$ has to be 
imposed to ensure the vanishing of the boundary terms. The 
condition can be understood easily as the residual symmetry of the string worldsheet is just a constant shift in the
coordinates and there is no analogous APD gauge transformation 
The second one implies that there are no $R\otimes R$
one forms on the boundary string theory. In case of the membrane 
ending on 9-branes this is easy to understand as
the boundary string theory is heterotic string theory.

Similarly linear supersymmetry transformations can be performed on the
rest of the vertex operators to see that they respect the algebra. Details
can be found in \cite{dnp}. The non-linear supersymmetry transformations
are more complicated and difficult to check, and an attempt in that
direction is given in \cite{dnp}. However, our vertex operators are further
confirmed by a series of other consistency checks, namely their invariance
under space-time gauge transformations and double dimensional reduction to yield superstring vertex operators. 
in \cite{dnp}. To check for the invariance under $\delta h_{ab}= k_{(a }\xi_{  b)}$ of the graviton vertex, we find
\begin{eqnarray}
\delta V_{\xi}&= &\left[D(k\cdot X) \left(D(\xi\cdot X) - R^{ab}\xi_ak_b\right) - \{k\cdot X, X^c\}\{\xi\cdot X,X^c\} \right.+ \nn \\
&&\left. 3\{k\cdot X,X^c\}R^{abc}\xi_ak_b - \frac{i }2\th\xi_a \g^a\{k\cdot X,\th\} - \frac{i }2\th \{\xi\cdot X,\th\}\right]e^{-i  k\cdot X}
\end{eqnarray}
By partially integrating the first two terms, and using the equations of motion for $X^a$ and $\theta$ (\ref{xeq},\ref{theq}), the terms completely cancel each other. Note however for the open super membrane, at the boundaries we have to impose $k_m\xi_M=0$, which is perfectly
consistent with the earlier condition $h_{mM}=0$. For details of the 
other vertex operators, we refer to \cite{dnp}.
We review the reduction to superstring vertex operators for the
graviton briefly here (we denote the string coordiantes as $X^i$). 
The double dimensional reduction consists of the identification of 
one of the supermembrane directions say $X^9=\sigma^2$ as compact, 
to get a string in 10 dimensions. In the first approximation
this means that $X^{i}$ are independent of $\sigma^2$, and hence 
$\{X^a,X^b\}\neq0$ only when $a$ or $b = 9$. Thus:
\begin{equation}
\{X^a,X^b\}= \partial_1 X^{[a}\delta^{b]}_9 
\label{ddr1}
\end{equation}
The SO(9) spinors also decompose into two SO(8) spinors $(S_{\a}, S_{\dot\a})$.
Thus ($i,j=1,..,8$ and $\G^i$ are SO(8) matrices):
\bea
R^{ij}  &=& \ft14 S \G^{ij} S  + \ft14 \tS \G^{ij} \tS  \qquad \qquad
            R^{i9} = \ft12 \tS \G^i S     \label{ddr2} \\
R^{ijk} &=& \ft16 S \G^{ijk} \tS   \qquad \qquad
    R^{ij9} = \ft1{12} S \G^{ij} S - \ft1{12} \tS \G^{ij} \tS \label{ddr3}
\eea

Using the above in the graviton vertex, one finds that the $h_{ij}$ should
give the $NS\otimes NS$ vertex operator of type IIA superstring, while
the $h_{i9}$ should give the $R\otimes R$ one form vertex in 10 dimenions.
Despite the apparent differences in the structure of the vertex operators,
we recover them perfectly starting from (\ref{Vh}) and using (\ref{ddr1},\ref{ddr2},\ref{ddr3}). 
The vertex operator in 10 dimensions is:
\bea
(V_h)_{DDR} &=& h_{ij} \le[ \partial_0 X^i \partial_0 X^j -\partial_1 X^i \partial_1 X^j
           - \ft12 \partial_0 X^i (S\G^{jm} S + \tS \G^{jm} \tS ) k_m  \re.\nn \\
   && \le.\qquad + \ft12 \partial_1 X^i (S\G^{jm} S - \tS \G^{jm} \tS ) k_m
   + \ft14 S \G^{jm} S \, \tS \G^{jn} \tS k_m k_n \re] e^{-i k\cdot X} \nn \\
  &=& h_{ij} \le( \partial_+ X^i - \ft12 S\G^{im} S k_m \re)
        \le( \partial_- X^j - \ft12 \tS\G^{in} \tS k_n \re) e^{-\i k\cdot X}
\eea

For the $h_{i9}$, we get:
\be
h_{i9} \le[ -i\th \g^i \partial_1 \th + 2 \partial_0 X^i R^{m9} k_m
  + \partial_1 X^j R^{ijm} k_m +2 R^{im} R^{9n} k_m k_n \re] e^{-\i k\cdot X}
\ee
Again the quartic terms are easily seen to agree. To get rid of
the derivatives on $\th$, which are absent in the superstring
vertices, we make use of the superstring equations of motion
$\partial_1 S = \partial_0 S$ and $\partial_1 \tS = - \partial_0 \tS$, and integrate the
resulting expression by parts to find

\be
k_{i} h_{j9} \le[ S \G^{ij} \G^{k} \tS \, \partial_- X^k -
       S \G^{k} \G^{ij} \tS \, \partial_+ X^k \re]  e^{-\i k\cdot X}
\ee

The same can be repeated for the three form vertex operator and the
gravitino vertex operators. Thus, we have the complete supermembrane
vertex operators both for closed as well as open supermembranes.
It should be mentioned that in the case of open membranes ending on
9-branes, heterotic string vertex operators can be recovered on the
boundary using the same techniques of double dimensional reduction.
Except, the additional massless states of the $E_8\times E_8$
gauge fields are not there. It shall be an interesting exercise 
to look for them in the open supermembrane spectrum. Also, the dimensional
reduction for the open supermembrane leads
to Type I' theory \cite{het}. This is
a manifestation of the web of dualities relating the various string theories.
\subsection{Matrix theory vertex operators}
Since the Matrix-regularisation of the supermembrane
is a straightforward procedure described in section 2.2.
the vertex operators can be easily applied to Matrix theory.
The coordinates $X,\theta$ transform in the adjoint of
SU(N) and hence are $(N^2-1) \times (N^2-1)$ matrices (for the closed membrane).
The continuum integral is replaced by a Trace operation.
Hence the graviton vertex (written in configuration space)
shall be of the form:
\begin{eqnarray}
{\bf V_{h}} &= &Tr \left[ \left\{ \dot{\bf X}\dot {\bf X} + [{\bf X^a},{\bf X^b}]^2 + {\Th\g^a}\left[{\bf X}^b,\Th\right]
-2\dot {\bf X^a} {\bf R}^{bc} \frac{\partial}{\partial{\bf X^c}} \right.\right.\nn\\&& \left.\left. - 6 \left[{\bf X}^a,{\bf X}^c\right]{\bf R}^{bcd}\frac{\partial}{\partial {\bf X^d}} + 2(\Th\g^{ac}\Th)(\Th\g^{bd}\Th)\frac{\partial}{\partial {\bf X}^c}\frac{\partial}{\partial{\bf X}^d}\right\}h_{ab}(\bf X)\right]
\nn \\
\end{eqnarray}
This agrees with previous calculations of linearised Matrix current in arbitrary backgrounds determined up to $O(\th^2)$ \cite{tay1}.
Note that our vertex operators are known to all orders in $\th$ and unlike as expected (i.e. terms up to $\th^{32}$), they contain terms only up to $O(\th^5)$. Note for the open-membrane also we can suitably regularise
remembering to replace the Dirichlet direction by SO(N) adjoint matrices and the rest by symmetric
traceless ones.
It remains now to implement the above in a scattering
amplitude calculation.

\section{Scattering Amplitudes}

We now turn to the discussion of three point tree level scattering amplitudes.
For this it is advantageous to work in the framework of the 
finite $N$ matrix theory. In order to define a tree level amplitude we
first split off the center of mass degrees of freedom of the matrices by
writing
\be
{\bf X}^a={x^a}\, \mathbb{1} + \bX^a \qquad
{\bf \Theta}^a={\theta^a}\, \mathbb{1} + \hat{\bf \Theta}^a 
\label{zero}
\ee
with traceless matrices $\hat{\bf X}^a$ and $\hat{\bf \Theta}$. An 
asymptotic 1-graviton state in matrix theory is then given by
\be
|\mbox{IN}\rangle\rangle= 
|k_1,h_1\rangle_{{x,\theta}} \otimes 
|\mbox{GS}\rangle_{\hat{\bf X},\hat\Theta}
^{\mbox{\tiny SU(N)}}
\ee
where $|k_1,h_1\rangle$ is the graviton state of the superparticle
\cite{pw,ggk} and 
$|\mbox{GS}\rangle$ denotes the {\it exact} SU($N$) normalized zero energy
groundstate, whose explicit form is unknown but is believed to exist \cite{gs}. 
The tree level three point amplitude is then defined by
\be
{\cal A}_{\mbox{\tiny 3-point}}=\langle\langle 
\,1\,|\, V_2\,|\,3\,\rangle\rangle
\label{3pt}
\ee
where one inserts the graviton vertex operator
\be
V_2=h^{(2)}_{ab}\, \ft 1 N \, \mbox{STr}\Bigl[({p^a}\,{p^b}  + 2\, {p^a}\, 
\hat{\bf P}^b
+ \hat{\bf P}^a\,\hat{\bf P}^b+ [\bX^a,\bX^c]\, [\bX^b,\bX^c]\, )\,
e^{ik\cdot\bX}\Bigr]\, e^{ik\cdot{x}}
+ \mbox{\small fermions}
\label{V2}
\ee
One may wonder how one could ever evaluate \eqn{3pt} upon inserting
\eqn{V2} without the knowledge
of $|\mbox{GS}\rangle$. The first contribution to \eqn{3pt} takes the
form
\be
\langle k_1,h_1| {p^a}\,{p^b}e^{ik\cdot{x}}|k_3,h_3\rangle\,
h^{(2)}_{ab}\,  
\langle\mbox{GS}|\mbox{STr}\, e^{ik\cdot\bX}|\mbox{GS}\rangle
\label{kh}
\ee
Now by SO(9) covariance 
$\langle\mbox{GS}|\mbox{STr}\, e^{ik\cdot\bX}|\mbox{GS}\rangle=N$, as
the only SO(9) scalar it could depend on would be $k^2$ which vanishes
on shell. It must then be a constant which is fixed to be $N$ by considering the
$k^a\rightarrow 0$ limit. Remarkably the remaining two terms in \eqn{3pt}
upon inserting \eqn{V2} vanish by a combination of SO(9) covariance
and on-shell arguments:
\be
h_{ab} \, \langle\mbox{GS}|\mbox{STr}\, \hat{\bf P^b} \, e^{ik\cdot\bX}
|\mbox{GS}\rangle \sim k^b\, h_{ab}=0
\ee
$$
h_{ab} \, \langle\mbox{GS}|\mbox{STr}\Bigl[( \hat{\bf P^a}\, \hat{\bf P^b} 
+ [\bX^a,\bX^c]\, [\bX^b,\bX^c]\, )
\, e^{ik\cdot\bX}\Bigl ]
|\mbox{GS}\rangle \sim (k^a\, k^b+ c\, \delta^{ab})\, h_{ab}=0
$$
But as the first correlator in \eqn{kh} is nothing but the bosonic
contribution to the 3-point
$d=11$ {\it superparticle} amplitude \cite{ggk} and as the fermionic terms
work out in a similar fashion we see that our 3-point tree level
amplitude 
\be 
\langle\langle \,1\,|\, V_2\,|\,3\,\rangle\rangle= 
\langle \,1\,|\, V_2\,|\,3\,\rangle_{{x,\theta}}\, \langle
\mbox{GS}|\mbox{GS}\rangle
\ee
agrees with the 3-point amplitude of $d=11$ supergravity!

Clearly the next step would be to study $n$-point tree level amplitudes
which should be given by
\be
{\cal A}_{\mbox{\tiny n-point}}
=\langle\langle \,1\,|\, V_2\, \Delta\, V_3\, \Delta \ldots \Delta\, V_{n-1}
|\,n\,\rangle\rangle
\label{npt}
\ee
where $\Delta$ denotes the propagator $1/(\ft 1 2\, {p_0}^2 + \hat{\bf H})$
built from the {\it interacting} membrane Hamiltonian $\hat{\bf H}$.
However, now we expect the details of the groundstate $|\mbox{GS}\rangle$
to enter the computation. Developing some perturbative scheme for
calculating \eqn{npt} would be highly desirable, but is conceivably
very complicated as it must involve an expansion in both the propagator
$\Delta$ and the groundstate $|\mbox{GS}\rangle$. In other words, even
though we now have the vertex operators for the massless states, we
are not able with present techniques to calculate the (super)membrane
analogue of the Veneziano amplitude. The complications are easy to 
understand from the figure: whereas for the string, the intermediate 
states are just the massive string states, and therefore completely under
control, we do not know how to interpret (and to manipulate) the
states that propagate between two vertices in the case of the membrane.
On the other hand, if we could master this calculation we might be
able to obtain the full result to all orders ``in one go'': there would
be no need for unitarity corrections and the like in this membrane
amplitude!

\begin{figure}[t]
\begin{center}
\begin{tabular}{ccc}
\epsfxsize=5cm
 \epsfbox{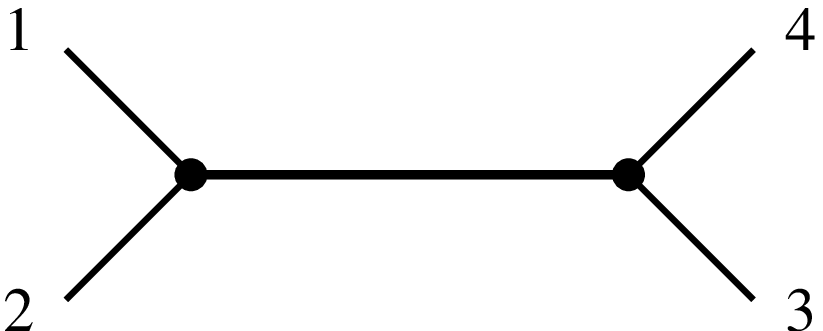} & &
\epsfxsize=5cm\epsfbox{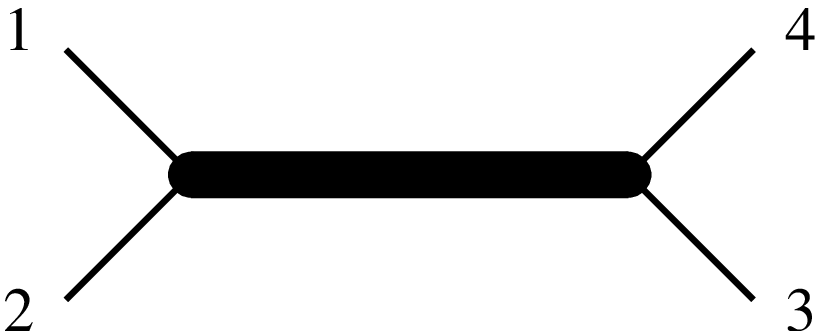}\cr 
 $H\sim P^2+X'^2$ & & $H\sim P^2+\{X,X\}^2$
\end{tabular}
\end{center}
\caption{The Veneziano amplitude for Strings and Membranes}
\end{figure}

Instead we shall briefly comment on attempts to compute loop
amplitudes within this scenario. Here, led by the formalism in
light cone superstring and superparticle theory, we propose to
define a membrane  $n$-point one-loop amplitude by the expression
\be
{\cal A}_{\mbox{ \tiny 1-loop, n-point}}
=\int d^{11}{p_0}\, \mbox{\bf Tr} ( \Delta\, V_1 \,
\Delta\, V_2\,  \Delta\, V_3\,  \Delta\, V_4\,  \ldots \Delta \, V_n)
\label{1lp}
\ee
where the trace is over the Hilbert space of $\hat{\bf H}$. Again
this appears as a daunting task, however the zero mode sector of
\eqn{1lp} already yields some amount of information. In particular
the trace over the fermionic zero mode $\theta$ of \eqn{zero}
tells us that all 2 and 3 particle amplitudes vanish at one loop,
as at least four vertex operators ($\leq$ 16 $\theta$'s) 
are needed to saturate the
fermion zero mode trace
\be
\mbox{\bf Tr} (\theta_{\alpha_1}\ldots
\theta_{\alpha_{{N}}})_{{\theta}} = \delta_{{N},16}\,
\epsilon^{\alpha_1\ldots\alpha_{16}}
\ee
In the pure graviton sector the first non-vanishing amplitude is
then the 4-graviton amplitude whose leading momentum dependence
is given by
\be
{\cal A}_{4h} = \epsilon^{\alpha_1\ldots\alpha_{16}}\, 
\gamma_{\alpha_1\alpha_2}^{a_1\, a_2}\ldots 
\gamma_{\alpha_{15}\alpha_{16}}^{a_{15}\, a_{16}}\,\,
R^{(1)}_{a_1a_2a_3a_4}\ldots R^{(4)}_{a_{13}a_{14}a_{15}a_{16}}
\int d^{11}\, {p_0}\, \mbox{\bf Tr}^\prime\, \Delta^4 \, .
\ee
We thus see the emergence of the expected ${\cal R}^4$ term 
\cite{gg,pierre} in the kinematical sector. For the remaining trace it
is useful to introduce a Schwinger time parametrisation of the propagator
$\Delta=\int_0^\infty dt \exp[-t(p_0^2/2+\hat{\bf H})]$. Then the remaining
trace factorizes into the product of the {\it classical} bosonic 
partition function and the interacting contribution of the quantum 
fluctuations. For an evaluation of the classical contribution to this
trace see \cite{bnpw}.
\vspace{.5cm}

\noindent
{\bf Note added:}
\noindent
After this paper was submitted for publication, progress has been made 
in reduction of supermembrane to Matrix string theory \cite{seki}, determination of Matrix theory currents \cite{rams} and towards covariant quantisation of the bosonic \cite{haya} and supersymmetric \cite{berk} membrane.


\begin{thebibliography}{99}
%
\bibitem{bst} E. Bergshoeff, E. Sezgin and P.K. Townsend, \PLB{189}{1987}{75};
 {Ann.\ Phys.} {\bf 185} (1988) 330. 
%
\bibitem{dwhn1} B. de Wit, J. Hoppe and H. Nicolai, {Nucl.\ Phys.} {\bf 305}
[FS23] (1988) 545.
\bibitem{dwhn2} B. de Wit, M. L\"uscher and H. Nicolai, \NPB{320}{1989}{135}.
\bibitem{sm} A.V.~Smilga, in \T{Supermembranes and Physics in
   2+1 Dimensions}, eds. M.J.~Duff, C.N.~Pope and E.~Sezgin,
   World Scientific (1990).
\bibitem{bfss} T.\ Banks, W.\ Fischler, S.H.\ Shenker and L.\
Susskind, {\it Phys. Rev.} {\bf D55} (1997) {5112}, \hepth{9610043}.
\bibitem{gs} A.V. Smilga, \NPB{266}{1986}{45}.
\\
P. Yi, Nucl. Phys. B{\bf 505} (1997) 307, {\tt hep-th/9704098}.
\\
S. Sethi, M. Stern,
Comm. Math. Phys. {\bf 194} (1998) 675, {\tt hep-th/9705046}.
\\
M. Porrati, A. Rozenberg,  Nucl. Phys. B{\bf 515} (1998) 184,
{\tt hep-th/9708119}.
\\
J. Fr\"ohlich and J. Hoppe, \CMP{191}{1998}{613}.
\\
M.B. Halpern and C. Schwartz, \IJMPA{13}{1998}{4367}.
\\
J. Fr\"ohlich, G.M. Graf, D. Hasler, J. Hoppe and S.T. Yau, \hepth{9904182}.
\\
G. Moore, N. Nekrasov and S. Shatashvili, \CMP{209}{2000}{77}.
\\
V.G. Kac and A.V. Smilga. \T{Normalized vacuum states in ${\cal N}=4$
supersymmetric Yang--Mills quantum mechanics with any gauge group},
\hepth{9908096}.
\\
J. Hoppe and J. Plefka, \T{The Asymptotic Groundstate of SU(3) Matrix theory},
\hepth{0002107}.
%
\bibitem{ggk} M.B. Green, H. Kwon and M. Gutperle, \JHEP{08}{1999}{012},
\hepth{9907155}.
%
\bibitem{pw} J. Plefka and A. Waldron, \NPB{512}{1998}{460}; 
\hepth{9710104}.
%
\bibitem{gg} M.B. Green, M. Gutperle and P. Vanhove,
      \PLB{409}{1997}{177}, \hepth{9706175}.
\bibitem{ssv}
M.B. Green and J.H. Schwarz, \PLB{136}{1984}{367}, \NPB{127}{1984}{285}.
\bibitem{dnp} A. Dasgupta, H. Nicolai and J. Plefka JHEP 05 (2000) 007, \hepth{0003}
\bibitem{pl} J. Plefka, \IJMPA{16}{2001}{660}
\bibitem{bnpw} B. Pioline, H. Nicolai, J. Plefka, A. Waldron \JHEP{0103}{2001}036.
\bibitem{hw} P. Horava E. Witten, \NPB{475}{1996}{94}, \NPB{460}{1996}{506}.
\bibitem{opm}D.S. Berman, P. Sundell \JHEP{0010}{2000}{014}\\ R. Gopakumar, S. Minwalla, N. Seiberg, A. Strominger \JHEP{0008}{2000}{008}
%
\bibitem{revs} M.J. Duff, \T{Supermembranes} \hepth{9611203}.\\
                  B. de Wit, \T{Supermembranes and Super Matrix Models},
  \hepth{9902051}.\\
           W. Taylor, \T{M(atrix) Theory: Matrix Quantum Mechanics as a Fundamental Theory} \hepth{0101126}

\bibitem{nh} H. Nicolai and R. Helling, \T{Supermembranes and M(atrix)
Theory}, \hepth{9809103}.
%
\bibitem{csj} E. Cremmer , B. Julia and J. Scherk, {Phys.\ Lett.}
     {\bf 76B} (1978) 409.
\bibitem{dpp} B. de Wit, K. Peeters and J. Plefka, \NPB{532}{1998}{99},
\hepth{9803209}.
\bibitem{tw} P. Pasti, M. Tonin \NPB{418}{1994}{337}\\
             I. Bandos, D. Sorokin, D. Volkov \PLB{352}{1995}{269}.
\bibitem{kg} M. T. Grisaru, M. E. Knutt, \PLB{500}{2001}{188}, \hepth{0011173}.
\bibitem{GH} J. Goldstone, unpublished.\\
J. Hoppe, in proc. Int. Workshop on Constraints Theory and Relativistic 
Dynamics, eds. G. Longhi and L. Lusanna, World Scientific (1987).
%
\bibitem{dwpp} B. de Wit, K. Peeters, J. Plefka, Nucl. Phys. Proc. Suppl 62 (1998) 405.
%
\bibitem{eza} K. Ezawa, Y. Matsuo, K. Murakami \PRD{57}{1998}{5118}.
\bibitem{ce} M. Cederwall, \MPLA{12}{1997}{2641}.
\bibitem{het} T. Banks, L. Motl \JHEP{9712}{1997}{004}, \hepth{9703218}\\
              D. Lowe \PLB{403}{1997}{243}, \hepth{9704041}\\
              S. Kachru, E. Silverstein \PLB{396}{1997}70, \hepth{9612162}\\
              U. H. Danielsson, G. Ferreti \IJMPA{12}{1997}4581, \hepth{9610082}
\bibitem{five} P. Townsend \PLB{373}{1996}{68}\\
               A. Strominger \PLB{383}{1996} 44.
\bibitem{mat} J. Hoppe, \PLB{215}{1988}{706}.\\
D. Fairlie, P. Fletcher and C.N. Zachos, \PLB{218}{1989}{203}.\\
B. de Wit, U. Marquard and H. Nicolai, {Comm. Math. Phys.}
    {\bf 128} (1990) 39.\\
M. Bordemann, E. Meinrenken and M. Schlichenmaier,
          \CMP{165}{1994}{281}, \hepth{9309134}.
\bibitem{rey} N. Kim, S-J. Rey \NPB{504}{1997}{189}, \hepth{9701139};
\bibitem{tay1} W. Taylor, M. V. Raamsdonk, \NPB{558}{1999}{63},\hepth{9904095}
\bibitem{pierre}
K.~Peeters, P.~Vanhove and A.~Westerberg,
\CQG{18}{2001}{843}, {{\tt hep-th/0010167}};\\
K.~Peeters, P.~Vanhove and A.~Westerberg,
\T{Supersymmetric $R^4$ actions and quantum corrections to superspace torsion constraints}
{{\tt hep-th/0010182}}.

\bibitem{seki} Y. ~Sekino, T.~ Yoneya, \NPB{619}{2001}{22}, \hepth{0108176}.
\bibitem{rams} M.~V. Raamsdonk, \T{Open Dielectric Branes}, \hepth{0112081}.
\bibitem{haya}M.~Hayakawa and N. Ishibashi, \NPB{614}{2001}{171}, \hepth{0107103}.
\bibitem{berk} N.~Berkovits, \T{Covariant Quantisation of the Supermembrane}, \hepth{0201151}.
\end{thebibliography}
\end{document}